\documentclass[11pt]{article}
\usepackage{amsfonts}
\usepackage{amssymb}
\usepackage{amsmath,amssymb}
\usepackage{bbm}

\def\slasha#1{\setbox0=\hbox{$#1$}#1\hskip-\wd0\hbox to\wd0{\hss\sl/\/\hss}}

\def\periodb#1{\setbox0=\hbox{$#1$}#1\hskip-\wd0\hbox to\wd0{-}}

\def\sfrac#1#2{{\textstyle\frac{#1}{#2}}}
\newcommand{\unit}{\mathbbm{1}}   

\newcommand{\frg}{\mathfrak{g}}
\newcommand{\frh}{\mathfrak{h}}
\newcommand{\CA}{\mathcal{A}}    
\newcommand{\CG}{\mathcal{G}} 
\newcommand{\CB}{\mathcal{B}} 
\newcommand{\CH}{\mathcal{H}} 
    
\newcommand{\CL}{\mathcal{L}}    
\newcommand{\CF}{\mathcal{F}}   
\newcommand{\CM}{\mathcal{M}}

\newcommand{\CE}{\mathcal{E}}    
\newcommand{\Ccal}{\mathcal{C}}

\newcommand{\R}{\mathbb{R}}     
\newcommand{\C}{\mathbb{C}}     
\newcommand{\Abb}{\mathbb{A}}     
\newcommand{\CPP}{{\mathbb{C}P}}    

\newcommand{\Id}{\mathrm{Id}} 
\newcommand{\dd}{\mathrm{d}}     
\newcommand{\dpar}{\partial}     
     
\newcommand{\dvol}{\mathrm{dvol}}

\newcommand{\tr}{\mathrm{tr}}     

\newcommand{\+}{\dagger}

\newcommand{\al}{{{\alpha}}} 
 \newcommand{\ga}{{{\gamma}}}

\newcommand{\veps}{{\varepsilon}} 
\newcommand{\vph}{{{\varphi}}} 
\newcommand{\Afla}{{{\mathcal A}^{\tt flat}}} 
\newcommand{\Gsm}{{{\mathcal G}^{}_{\tt small}}} 
\newcommand{\Gla}{{{\mathcal G}^{}_{\tt large}}} 

\makeatletter
\renewcommand*\l@section{\@dottedtocline{1}{1.5em}{4em}}
\@addtoreset{equation}{section}
\makeatother

\begin{document}
\begin{titlepage}
\setcounter{page}{0}
.
\vskip 3cm
\begin{center}
{\LARGE \bf Yang-Mills-Stueckelberg Theories, Framing}\\
\vskip 0.5cm
{\LARGE \bf and Local Breaking of Symmetries}
\vskip 1.5cm
{\Large Alexander D. Popov}
\vskip 1cm
{\em Institut f\"{u}r Theoretische Physik,
Leibniz Universit\"{a}t Hannover\\
Appelstra{\ss}e 2, 30167 Hannover, Germany}\\
{Email: alexander.popov@itp.uni-hannover.de}
\vskip 1.1cm
\end{center}
\begin{center}
{\bf Abstract}
\end{center}
We consider Yang-Mills theory with a compact structure group $G$ on a Lorentzian 4-manifold $M=\R\times\Sigma$ such that gauge transformations become identity on a submanifold $S$ of $\Sigma$ (framing over $S\subset\Sigma$). The space $S$ is not necessarily a boundary of $\Sigma$ and can have dimension $k{\le}3$. Framing of gauge bundles over  $S\subset\Sigma$ demands introduction of a $G$-valued function $\phi_S$ with support on $S$ and modification of Yang-Mills equations along $\R\times S\subset M$. The fields $\phi_S$ parametrize nonequivalent flat connections mapped into each other by a dynamical group $\CG_S$ changing gauge frames over $S$. It is shown that the charged condensate $\phi_S$ is the Stueckelberg field generating an effective mass of gluons in the domain $S$ of space $\Sigma$ and keeping them massless outside $S$. We argue that the local Stueckelberg field $\phi_S$ can be responsible for color confinement. We also briefly discuss local breaking of symmetries in gravity. It is shown that framing of the tangent bundle over a subspace of space-time makes gravitons massive in this subspace.

\end{titlepage}
\newpage
\setcounter{page}{1}

\section{Introduction}

\noindent
Let us consider gauge theory on the space-time manifold of the form $M=\R\times\Sigma$, where $\R$ is the time axis and $\Sigma$ is the spacial hypersurface with the boundary $\dpar\Sigma$. Such $M$ has the time-like boundary $\dpar M=\R\times\dpar\Sigma$ and the variation of action functional will have a boundary term. In conventional field theory one usually assumes that the boundary term can be dropped after imposing proper boundary conditions. However, this is not always possible without losing essential information, e.g. for dyon in $\Sigma = \R^3$ with the spacial boundary
$S^2_\infty =\dpar\Sigma$. In~\cite{GSW, BCT} it was proposed to circumvent this difficulty by introducing additional dynamical variables on the boundary $\dpar\Sigma$ with an appropriate action. The difference between what they call ``proper'' and ``improper'' gauge transformations was also discussed -- proper transformations do not change the physical state of the system where improper ones do. Now they are often called ``small'' and ``large'' gauge transformations.

In recent years, the above ideas were developed and applied to various new problems in gauge theories, e.g. to asymptotic symmetries~\cite{Str1}, entanglement entropy and edge modes~\cite{DF, BMV}, soft theorems~\cite{Str2} and many others. The additional variables in gauge theories are defined on boundaries (asymptotic or at finite distance) of space-time and they are often referred as edge modes. Action functionals and dynamics of boundary variables $\vph$ were discussed in~\cite{GSW, BCT} and recently e.g. in~\cite{DF, BMV}. Space-time is divided into the bulk and a boundary and variation of the action gives the standard Yang-Mills equations in the bulk and some equations for $\vph$ on the boundary. However, there are still many uncertainties in the interpretation of edge modes $\vph$ and in the choice of action for them.

Equating matrices of gauge transformations  on the boundary to unity means imposing the Dirichlet boundary conditions, and this is equivalent to {\it framing} the gauge bundle $E\to M$ over the boundary $\dpar M$ (see e.g.~\cite{Do1}). Framing means a choice of an ordered basis in fibres of the gauge bundle $E$ and this choice is parametrized by boundary fields  $\vph$.  In physics literature this idea goes back to~\cite{RT} (for gravity) and~\cite{GSW} (for gauge theories), where it was proposed to extend the phase space by boundary variables.

The key word for understanding asymptotic symmetries and edge modes is {\it framing}. Note that framing of the bundle $E\to M$ can be introduced not only on the boundary of the manifold $M$, but on any subspace of $M$. For example, when studying instantons, 
Donaldson introduced framing over the point
$\{\infty\}$ in $S^4=\R^4\cup\{\infty\}$ and over the subspace $\CPP^1$ in $\CPP^2$~\cite{Do2}. More general cases were considered e.g. in~\cite{Lu}. However, Yang-Mills equations for connections on framed bundles as well as auxiliary fields and action functionals for them were not discussed in the mathematical literature. In this paper we will focus on these issues by considering space-time $M=\R\times\Sigma$ and framing of the gauge bundle $E$ over a 3-dimensional submanifold $S$ in $\Sigma$.

We will consider a principal $G$-bundle $P$ over $M=\R\times\Sigma$ and the associated Hermitian vector bundle $E\to M$, where $G$ is a compact structure group. Framing of $E$ over $S\subset\Sigma$ for any $t\in\R$ breaks gauge symmetry since allowed gauge transformations have to be identity on $\R\times S$. We show that gauge invariance can be restored after introducing a $G$-valued field $\phi_S$ on $S$ which parametrizes the longitudinal component $\CA^L$ of a connection $\CA$ on the bundle $E$. These {\it framing fields} $\phi_S$ form an infinite-dimensional group $\CG_S$ which is a subgroup of the automorphism group of $E$. The Lie algebra Lie$\,\CG_S$ is parametrized by fields $\pi_S\in\,$ Lie$\,\CG_S$ on $S$ which are conjugate momenta of $\phi_S$ after identification of Lie$\,\CG_S$ with the dual space (Lie$\,\CG_S$)$^*$. The gauge invariance is restored on the phase space of Yang-Mills theory extended  by the conjugate pairs of variables $(\phi_S, \pi_S)$ from the cotangent bundle $T^*\CG_S$ of $\CG_S$. We show that $\phi_S$ is the Stueckelberg field supported on $S$ and generating a mass $m_S$ of gluons which is nonzero only on $S\subset\Sigma$. Note also that $\phi_S$ parametrizes the non-trivial flat connection $\Afla = \phi_S^{-1}\dd\phi_S$ inside $S$ which becomes trivial $\Afla = 0$ outside $S$. Thus, framing of the gauge bundle $E$ over $S$ creates a bubble of gluon condensate with positive energy density preventing the expansion of the bubble  with quarks. After discussing local breaking of gauge symmetry, we introduce the local Stueckelberg fields for diffeomorphisms and Lorentz transformations and briefly discuss local breaking of symmetries in gravity.

\section{Fibre bundles and automorphisms}

\noindent
{\bf $G$-bundles.}  Let $M$ be a smooth Lorentzian four-manifold, $G$ a compact semisimple Lie group, $\frg$ its Lie algebra and $P=P(M,G)$ a principal $G$-bundle over $M$. In gauge theories one also considers three bundles asociated with $P$:
\begin{itemize}
\item the bundle of groups Int$\,P=P\times_GG$, where $G$ acts on itself by internal automorphisms, $h\mapsto ghg^{-1}$ for $h, g\in G$,
\item the bundle of Lie algebras Ad$\,P=P\times_G\frg$, where $G$ acts on $\frg$ by adjoint action, $a\mapsto gag^{-1}$ for $a\in \frg$, $g\in G$,
\item a complex vector bundle $E=P\times_GV$, where $V$ is the space of some irreducible representation of $G$.
\end{itemize}
We will consider complex vector bundles $E\to M$ of rank $n$ with the vector space $V=\C^n$ endowed with a Hermitian metric, i.e. $G$ can be considered as a closed subgroup of SU$(n)$.

\smallskip

\noindent
{\bf Frames.} The bundles $P$, $E$, Int$\,P$ and Ad$\,P$ are closely related to each other. By definition, $E$ is associated with $P$ and $P$ is the bundle of $G$-frames on $E$. The fibre of $P\to M$ over a point $x\in M$ is the group $G_x$ of all ordered bases, or {\it frames}, for the fibre $E_x\cong\C^n$ of the vector bundle $E\to M$. Sections of $E$ are $\C^n$-valued functions on $M$, they are matter fields (quarks after tensoring with spinor bundle).

\smallskip

\noindent
{\bf Automorphisms.} Global sections of the bundle Int$\,P$ are used for defining change of frames on the vector bundle $E\to M$. We consider smooth sections of Int$\,P$. They form an infinite-dimensional group $\CG = \Gamma^\infty (M$, Int$\,P$) whose infinitesimal action on $E$ is defined by smooth sections of the bundle  Ad$\,P$, i.e. Lie$\,\CG = \Gamma^\infty(M$, Ad$\,P)$. Elements of $\CG$ are {\it automorphisms} of $P$ inducing identity transformations of $M$, i.e. $\Gamma^\infty (M$, Int$\,P)\subset\,$Aut$\,P$. The group $\CG$ is also the group of $G$-automorphisms of $E$ covering the identity map of $M$, i.e. $\CG\cong\,$Aut$_GE$.

\smallskip

\noindent
{\bf Connections.} Let $\CA$ be a connection one-form on the principal bundle $P(M,G)\to M$, and $\CF = \dd\CA + \CA\wedge\CA$ its curvature. Associated bundles Int$\,P$, Ad$\,P$ and $E$ inherit their connection $\CA$ from the bundle $P$. 

We denote by $\Abb$ the space of connections on $P$. This is also the space of connections on $E$ and on any other bundles associated with $P$. The infinite-dimensional group $\CG$ acts on $\Abb$ by the standard formula
\begin{equation}\label{2.1}
\CA\mapsto\CA^g=g^{-1}\CA g+ g^{-1}\dd g\ ,
\end{equation}
for $g\in \CG = \Gamma^\infty (M$, Int$\,P$). The infinitesimal action of automorphisms is defined by formula
\begin{equation}\label{2.2}
\CA\mapsto\delta_\xi\CA=\dd_{\CA}\xi=\dd\xi + [\CA , \xi]\ ,
\end{equation}
where $\xi\in\,$Lie$\,\CG = \Gamma^\infty(M$, Ad$\,P)$.

 \smallskip

\noindent
{\bf Remark.} One often considers the subgroup $\CG^{x_0}$ of $\CG$ which consists from those $g\in\CG$ which are unity at some point $x_0\in M$. Later we will see that this corresponds to framing of the bundle $P$ over $x_0$. The group $\CG^{x_0}$ of pointed automorphisms acts {\it freely} on the space $\Abb$ of connections that is important in many cases. Obviuosly, the quotient space is the group $G=\CG/\CG^{x_0}$. 

\smallskip

\noindent
{\bf Gauge transformations.} All automorphisms from the group $\CG = \Gamma^\infty (M$, Int$\,P$) map the bundle $E$ into itself. They change frames in $E$ and transform the connection $\CA$ on $E$ by the formula  \eqref{2.1} with $g\in\CG$. In any particular model one
 imposes on $E$  (explicitly or implicitly) the restriction that its frame is fixed somewhere, most often over the boundary of $M$ or at a point $x_0\in M$. Only a subgroup of $\CG$, let us denote it $\Gsm$, will preserve this condition. The subgroup $\Gsm$ is a normal subgroup of $\CG$, it means that
\begin{equation}\label{2.3}
gng^{-1}\in\Gsm
\end{equation}
for any $n\in\Gsm$ and $g\in\CG$. This is why the quotient,
\begin{equation}\label{2.4}
{\Gla} :=\CG /\Gsm\ ,
\end{equation}
is also a group and the group $\CG$ can be presented in the form of the semi-direct product
\begin{equation}\label{2.5}
{\CG} =\Gsm\rtimes\Gla
\end{equation}
of the group $\Gsm$ and $\Gla$.

Let us denote by $N$ the submanifold of $M$ over which the frames of $E$ are fixed. The pair
\begin{equation}\label{2.6}
\Bigl({\rm bundle}\  E\to M,\ {\rm frames\ over}\ N\subset M\ {\rm are\ fixed}\Bigr)
\end{equation}
is called {\it framed bundle}. The group $\Gsm$ preserves framing, maps the framed bundle into itself and it is considered as the group of
gauge transformations describing {\it redundancy}. Therefore, connections $\CA\in\Abb$ on framed bundles $E$ related by the transformations \eqref{2.1} for $g$ from $\Gsm$ have to be considered as equivalent. The {\it moduli space} of such connections is defined as the quotient
\begin{equation}\label{2.7}
\CM = \Abb / \Gsm
\end{equation}
since $\Gsm$ defines redundancy of description.

\smallskip

\noindent
{\bf Physical symmetries.} The pairs \eqref{2.6} with different frames over $N\subset M$ are considered as ``physically'' different. The group $\Gla$ defined in \eqref{2.4} does not preserve the framed bundle \eqref{2.6} and maps it into another nonequivalent bundle. Hence $\Gla$ is a dynamical group transforming the connection $\CA\in\Abb$ into some nonequivalent $\CA^g$ for $g\in \Gla$. This nonequivalence can be observed in many particular models (see e.g.~\cite{Str1, DF, BMV, Str2, Do1}). Note that acting by $\Gla$ on the perturbative vacuum $\CA=0$, we obtain flat connections
\begin{equation}\label{2.8}
\CA=g^{-1}\dd g
\end{equation}
parametrized by the infinite-dimensional space $\Gla/G$.

For framing over boundaries, fields parametrized by $\Gla$ are considered as ``edge modes''. The group $\CG$ acts on the moduli space \eqref{2.7} with the stability subgroup $\Gsm$ and we can define the fibration
\begin{equation}\label{2.9}
p:\quad \CM = \Abb / \Gsm\quad\stackrel{\Gla}{\longrightarrow}\quad\Abb /\CG =:\CM_0
\end{equation}
with the group $\Gla$ as fibres. The moduli space $\CM_0$ parametrizes $\Gla$-invariant connections. 

\section{Yang-Mills theory in the bulk}

\noindent
{\bf Manifold $M=\R\times\Sigma$.} The discussion in Section 2 is general and applied to any manifold $M$ and a bundle $E\to M$ framed over a submanifold $N$ of $M$ of any dimension dim$\,N\le 4$. In this paper we will consider 4-manifolds $M=\R\times\Sigma$ with the boundary $\dpar M=\R\times\dpar\Sigma$, where $\dpar\Sigma$ can be at infinity, e.g. the two-sphere $S^2_\infty =\dpar\R^3$ at infinity in $\Sigma = \R^3$. We want to consider bundles $E$ over $\R\times\Sigma$ framed over a submanifold $\R\times S$, where $S$ is a 3-dimensional submanifold of finite volume in $\Sigma$. We restrict all fields to be trivial on the spatial boundary $\dpar\Sigma$ since we want to study effects of framing over $S\subset\Sigma$ in the bulk not clouded by boundary contributions.

\smallskip

\noindent
{\bf Bundles.} Consider the restriction of the bundle $P\to\R\times\Sigma$ to $\Sigma_t = \{t\}\times\Sigma$ and denote it $P_{\Sigma_t}$. Then we have a family $\{P_{\Sigma_t}\}$ of bundles $P_{\Sigma_t}\to \Sigma_t$ and all of them are isomorphic to the bundle $P_{\Sigma}\to \Sigma$ at $t=0$. The dynamics of Yang-Mills theory on $M=\R\times\Sigma$ are governed by ``paths'' in a moduli space of connections on the principal bundle $P_\Sigma$ over $\Sigma$, which we choose to be trivial $P_\Sigma = \Sigma\times G$. Similarly, the complex vector bundle $E=M\times \C^n$  is equivalent to the family $\{E_{\Sigma_t}\}$ of bundles $E_{\Sigma_t}$, each of which is isomorphic to the bundle $E_\Sigma =\Sigma\times\C^n$.

We will consider the automorphism group of bundles $P$ and $E$ restricted to the group
\begin{equation}\label{3.1}
\CG = \{g\in C^\infty (\R\times\Sigma , G)\mid g^{}_{|\dpar M} = \Id\}\ .
\end{equation}
This group is the same as the family $\{\CG_{\Sigma_t}\}$ of groups $\CG_{\Sigma_t}$ isomorphic to the group
\begin{equation}\label{3.2}
\CG_\Sigma = \{g\in C^\infty (\Sigma , G)\mid g^{}_{|\dpar\Sigma} = \Id\}\ .
\end{equation}
On $M=\R\times\Sigma$ we introduce the metric
\begin{equation}\label{3.3}
\dd s^2 = -\dd t^2 + \dd s^2_\Sigma = -\dd t^2 + \delta_{ab}e^ae^b\ ,
\end{equation}
where $\{e^a\}, a=1,2,3$, is an orthonormal basis of one-forms on $\Sigma$. Since we consider matrix groups $G$, the metric on $\frg =\,$Lie$\,G$ is defined by trace $\tr$, and so the metrics on $\Sigma$ and on $\frg$ induce the natural inner product on Lie$\,\CG_\Sigma$ defined by
\begin{equation}\label{3.4}
\langle \xi_1, \xi_2\rangle = -\int_\Sigma\dvol^{}_\Sigma\,\tr (\xi_1\xi_2)\ ,
\end{equation}
for $\xi_1, \xi_2\in C^\infty (\Sigma, \frg )$ and $\dvol^{}_\Sigma =e^1\wedge e^2\wedge e^3$.

\smallskip

\noindent
{\bf Action functional.} The gauge potential $\CA$ for a connection on the bundles $P$ and $E$ can be written as
\begin{equation}\label{3.5}
\CA = \CA_\mu e^\mu = \CA_t\dd t + \CA_a e^a
\end{equation}
and for the curvature two-form $\CF$ we have
\begin{equation}\label{3.6}
\CF =\sfrac12\CF_{\mu\nu} e^\mu\wedge e^\nu = \CF_{ta}\dd t\wedge e^a + \sfrac12\CF_{ab}e^a\wedge e^b\ ,
\end{equation}
where $\mu = 0,1,2,3,\ e^0:=\dd t$ and $e^a$ are introduced in \eqref{3.3}. 

The Yang-Mills action functional is
\begin{equation}\label{3.7}
S{=}\int_{\R\times\Sigma}\!\dd t\wedge\dvol^{}_\Sigma\CL{=}\sfrac14\int_{\R\times\Sigma}\!\dd t\wedge\dvol^{}_\Sigma\tr (\CF_{\mu\nu}\CF^{\mu\nu}){=}-\sfrac12\int_{\R\times\Sigma}\dd t\wedge\dvol^{}_\Sigma\tr(\CF_{ta}\CF_{ta}{-}\sfrac12\CF_{ab}\CF_{ab})\ .
\end{equation}
The Lagrangian density reads
\begin{equation}\label{3.8}
\CL = -\sfrac12\tr (\CE_a\CE_b - \CB_a\CB_a )\ ,
\end{equation}
where 
\begin{equation}\label{3.9}
\CE_a:=\CF_{ta}\quad{\rm and}\quad\CB_a:=\sfrac12\veps_{abc}\CF_{bc}\ .
\end{equation}
The action \eqref{3.7} contains ``coordinates'' $\CA_t$, $\CA_a$ from the space $\Abb$ of all connections and ``velocities'' $\dot\CA_a=\dpar_t\CA_a$. We see that $\CE_a=\dpar\CL /  \dpar\dot\CA_a$ from \eqref{3.9} are momenta conjugate to coordinates $\CA_a$ and $\CE_t=\dpar\CL /  \dpar\dot\CA_t=0$, i.e. $\CA_t$ is a nondynamical Lagrange multiplier.

The manifold $M=\R\times\Sigma$ has a time-like boundary $\dpar M=\R\times\dpar\Sigma$ and for nonempty $\dpar\Sigma$ the variation of the action \eqref{3.7} will contain the boundary term proportional to the integral of the 3-form
\begin{equation}\label{3.10}
\tr (\delta\CA\wedge \ast\CF)
\end{equation}
restricted to the boundary $\dpar M$. Here $\ast$ is the Hodge star operator on $M$. For studying framing for Yang-Mills in the bulk, we assume that $\CA$, $\CF$ vanish on $\dpar\Sigma$ and  $g\to\unit_n$ on $\dpar\Sigma$. Then we can rewrite the action functional \eqref{3.7} as the integral of the Lagrangian density
\begin{equation}\label{3.11}
\CL^\prime =-\tr (\CE_a\dot\CA_a - \CH + \CA_t\nabla_a\CE_a)\ ,
\end{equation}
where
\begin{equation}\label{3.12}
\CH = -\sfrac12\tr (\CE_a\CE_a + \CB_a\CB_a)
\end{equation}
is the Hamiltonian density.

\smallskip

\noindent
{\bf Gauss constraints.}  For vanishing boundary integral of \eqref{3.10}, the Euler-Lagrange equations following from \eqref{3.7} are the Gauss law constraint equation
\begin{equation}\label{3.13}
\nabla_a\CF_{ta} = 0\quad\Leftrightarrow\quad\nabla_a\CE_a=0
\end{equation}
and the dynamical equations
\begin{equation}\label{3.14}
\nabla_t\CF_{ta}+\nabla_b\CF_{ab} = 0\quad\Leftrightarrow\quad\nabla_t\CE_a+\veps_{abc}\nabla_b\CB_c=0\ ,
\end{equation}
where covariant derivatives contain not only gauge but also the Levi-Civita connection.

We already mentioned that the component $\CA_t$ is not a dynamical variable. Hence, it can be eliminated by using a transformation \eqref{2.1} generated by the group \eqref{3.1}. Then $\CA_t=0$, $\CE_a=\CF_{ta}=\dpar_t\CA_a$ and the group preserving this temporal gauge is the group \eqref{3.2}. This group acts on the bundles $P^{}_\Sigma$ and $E^{}_\Sigma$ with a connection $\CA=\CA_ae^a$. Thus we have the infinite-dimensional phase space $T^*\Abb^{}_\Sigma$ of pairs $(\CA_a, \CE_a)$ for a Hamiltonian formulation of Yang-Mills theory~\cite{FS}. This formulation uses the cotangent bundle $T^*\Abb^{}_\Sigma$ over the space $\Abb^{}_\Sigma$  of connections on the bundle $E^{}_\Sigma$ vanishing on $\dpar\Sigma$. 

The Gauss law equation  \eqref{3.13} is a time-independent constraint equation on variables $(\CA , \CE)\in T^*\Abb^{}_\Sigma$ defining a constraint surface 
\begin{equation}\label{3.15}
 \Ccal^{}_\Sigma\subset T^*\Abb^{}_\Sigma
\end{equation}
in the phase space $T^*\Abb^{}_\Sigma$. The group $\CG^{}_\Sigma$ maps this surface into itself and we can define the moduli space of pairs $(\CA, \CE )$,
\begin{equation}\label{3.16}
 \CM^{}_{\tt phs}=T^*\Abb^{}_\Sigma/\!/\CG^{}_\Sigma :=
 \Ccal^{}_\Sigma/\CG^{}_\Sigma\ ,
\end{equation}
as the quotient of $\Ccal^{}_\Sigma$ by the action of the remnant gauge group $\CG^{}_\Sigma$.

The double quotient notation in \eqref{3.16} refers to the symplectic reduction of the phase space $T^*\Abb^{}_\Sigma$ in two steps defined by 
 \eqref{3.15}  and \eqref{3.16}. After resolving constraints  \eqref{3.13}, six $\frg$-valued functions $(\CA_a, \CE_a)\in T^*\Abb^{}_\Sigma$ on $\Sigma$ reduce to four $\frg$-valued functions as independent variables parametrizing $\CM^{}_{\tt phs}$. In the Coulomb gauge $\nabla_a\CA_a=0$ one can split $(\CA , \CE)$, 
\begin{equation}\label{3.17}
\CA_a = \CA_a^T+\CA_a^L\quad\mbox{and}\quad\CE_a = \CE_a^T+\CE_a^L\ ,
\end{equation}
into {\it transverse} $(\CA^T, \CE^T)$ and {\it longitudinal} $(\CA^L, \CE^L)$ components. Then the pair $(\CA^T, \CE^T)$ is an element of the moduli space \eqref{3.16} and  $(\CA^L, \CE^L)$ are variables parametrizing the group $T^*\CG^{}_\Sigma$. After resolving the Gauss constraints \eqref{3.13}, the variables  $(\CA^L, \CE^L)$ are expressed via $(\CA^T, \CE^T)$~\cite{FS}. The group $T^*\CG^{}_\Sigma$ is the semidirect product
\begin{equation}\label{3.18}
\CG^{}_\Sigma\rtimes ({\rm Lie\,}\CG^{}_\Sigma)^*
\end{equation}
of the group $\CG^{}_\Sigma$ and the space $({\rm Lie\,}\CG^{}_\Sigma)^*$ dual to the Lie algebra ${\rm Lie\,}\CG^{}_\Sigma$ 
with the coadjoint action of $\CG^{}_\Sigma$ on $({\rm Lie\,}\CG^{}_\Sigma)^*$. We identify $({\rm Lie\,}\CG^{}_\Sigma)^*$ with ${\rm Lie\,}\CG^{}_\Sigma$ via the inner product
\eqref{3.4}.

The above picture follows from masslessness of $\CA$. Massless gluons have two polarization states (transverse). The third polarization state, the longitudinal one, is eliminated by the Gauss law constraints \eqref{3.13} and the remnant $\CG^{}_\Sigma$-invariance. In this case the reduction \eqref{3.16} can be defined via the projection
\begin{equation}\label{3.19}
\pi:\quad  T^*\Abb^{}_\Sigma\quad\stackrel{T^*\CG^{}_\Sigma}{\longrightarrow}\quad\CM^{}_{\tt phs}
\end{equation}
with the group $T^*\CG^{}_\Sigma$ as fibres. The group $T^*\CG^{}_\Sigma$ parametrizes the longitudinal variables $(\CA^L, \CE^L)$ of 
$(\CA, \CE)\in T^*\Abb^{}_\Sigma$. Finally, the dynamical equations \eqref{3.14} define a trajectory in moduli space $\CM^{}_{\tt phs}$.

\section{Framing in gauge theories}

\noindent
{\bf Framing.} From this point on, we consider Minkowski space $M=\R\times\Sigma$ with $\Sigma =\R^3$ and a compact (closed and bounded) embedded submanifold $S$ in $\Sigma$. Let $E^0_S$ be a {\it fixed} complex vector bundle over $S$,
\begin{equation}\label{4.1}
E^0_S=S\times\C^n\to S\ ,
\end{equation}
i.e. $E^0_S$ has a {\it fixed frame} in $E_x^0\cong\C^n$ for any $x\in S$. Consider a restriction of the complex vector bundle $E=\R\times\Sigma\times\C^n$ to $\R\times S$,
\begin{equation}\label{4.2}
E_S=E_{|S_t}\quad\mbox{for}\quad S_t=\{t\}\times S\subset\R\times S\ .
\end{equation}
The bundle $E$ is said to be framed over $S$ if for any $t\in\R$ there exist a map
\begin{equation}\label{4.3}
\phi_S: E_S\to E_S^0
\end{equation}
which defines an isomorphism of these bundles. The $G$-valued functions $\phi_S(t)$ are defined on $\Sigma =\R^3$ and {\it supported on} $S\subset\Sigma$.

\smallskip

\noindent
{\bf Remark.} The {\it support} of a complex-valued function $f$ on the space $\Sigma$ is defined as the smallest closed set $S$ containing all points $x\in\Sigma$ where $f(x)\ne 0$. The same notion is defined for $G$-valued functions $f{:}\,\Sigma\to G$ as the smallest closed set containing all points $x\in\Sigma$ for which $f(x){\ne}\,$Id. Functions with {\it compact support} are those whose closed support $S$ is a compact subset of $\Sigma$.

\smallskip

\noindent
{\bf Automorphisms of $(E, \phi_S)$.} Framed bundles are determined by the data
\begin{equation}\label{4.4}
(E,\ E_S^0,\ \phi_S: E_S\to E_S^0,\ S)
\end{equation}
which we denote for short $(E, \phi_S)$. We introduced the group \eqref{3.1} of automorphisms of the bundle $E$ which for any fixed $t\in\R$ is isomorphic to the group $\CG_\Sigma$ given in \eqref{3.2}. We are interested in a subgroup of $\CG$ preserving the pairs $(E, \phi_S)$. Obviously, the subgroup of $\CG$ preserving framing is
\begin{equation}\label{4.5}
\CG_0=\{g\in\CG\mid g^{}_{|\R\times S} = \Id\}
\end{equation}
and it can be considered as the group of gauge transformations of framed bundles $(E, \phi_S)$.

The group $\CG_0$ is a {\it normal subgroup} in $\CG$, see  \eqref{2.3} for $\Gsm =\CG_0$. Hence we can define coset space
\begin{equation}\label{4.6}
\tilde \CG_S= \CG/\CG_0\quad\Leftrightarrow\quad\CG =\CG_0\rtimes\tilde \CG_S\ .
\end{equation}
Elements of $\tilde \CG_S$ are equivalence classes
\begin{equation}\label{4.7}
g\CG_0 =\{gg_0\mid g_0\in\CG_0\}=\{g_0^\prime g\mid g_0^\prime\in\CG_0\}=\CG_0g
\end{equation}
for $g\in\CG$. They are multiplied as follows:
\begin{equation}\label{4.8}
(g\CG_0 )(f\CG_0)=(gf)\CG_0  \ .
\end{equation}
With this multiplication $\tilde \CG_S$ is the {\it quotient group}. There is a natural homomorphism
\begin{equation}\label{4.9}
q:\quad \CG\stackrel{\CG_0}{\longrightarrow}\tilde \CG_S
\end{equation}
given by $q(g)=g\CG_0$. It maps $\CG_0$ into the identity element of $\tilde\CG_S$, $\CG_0=\mbox{ker}\,q$.

To see what is the group $\tilde \CG_S$, we consider the group $\CG_S$ of $G$-valued functions supported on $S$ for any $t\in\R$. Then any element $\tilde g_S$ of $\tilde\CG_S$ can be represented as $\tilde g_S = g_Sg_0$,  where $g_0\in\CG_0$ and $g_S\in\CG_S$. In other words, $\tilde g_S\in\tilde\CG_S$ differs from $g_S\in\CG_S$ by a gauge transformation $g_0\in\CG_0$. Hence, we will use the group $\CG_S$ in our further discussions. Elements $\phi_S$ of this group are defined on the whole Minkowski space, $\phi_S\in\CG_S$ are time-dependent $G$-valued functions supported on $S\subset\Sigma =\R^3$. 

The group $\CG_S$ does not preserve framing and hence it is a dynamical group for gauge theories on framed bundles. In such theories the data  \eqref{4.4} are considered as observable data. Thus, for framed bundles the group \eqref{4.5} is the group $\Gsm$ and the group $\CG_S$ can be identified with the group $\Gla$ from Section 2. Both groups act on connections $\CA\in\Abb$ by the standard formula \eqref{2.1}.

\smallskip

\noindent
{\bf Longitudinal components.} Framing of the bundle $E\to\R\times\Sigma$ over $S\subset\Sigma$ for any time breaks the group $\CG$ of gauge transformations to its normal subgroup $\CG_0$ defined in \eqref{4.5}. The space $\Sigma =\R^3$ can be divided into two regions: 
$\Sigma\backslash S$ and $S$. Over $\Sigma\backslash S$, the group $\CG$  coincides with $\CG_0$ and hence we have the standard Yang-Mills model discussed in Section 3. Over $S$, automorphisms of the bundle $E$ are given by the group  $\CG_S\subset\CG$ which rotates frames of $E_S$ and hence $\CG_S$ is physical. In terms of \eqref{3.16} and \eqref{3.19}, the moduli space is
\begin{equation}\label{4.10}
T^*\Abb^{}_{\Sigma\backslash S}/\!/\CG^{}_{\Sigma\backslash S}\cong T^*(\Abb^{}_{\Sigma\backslash S}/\CG^{}_{\Sigma\backslash S})
\end{equation}
for fields at $x\in\Sigma\backslash S$ and it is $T^*\Abb^{}_S$ for $x\in S$ that follows from the definition
\begin{equation}\label{4.11}
\CM_{\tt phs}^{\tt framed} = T^*\Abb^{}_{\Sigma}/\!/\CG_0\ .
\end{equation}
This means that one cannot impose the Gauss law constraints  \eqref{3.13} and further $\CG_S$-invariance in the region $S\subset \Sigma$. This also means that in the decomposition  \eqref{3.17} the longitudinal part $(\CA^L, \CE^L)\in T^*\CG_S$ cannot be eliminated and we have three polarizations of gluons in $S$, i.e. they become effectively massive in the region $S\subset\Sigma$. Notice also that the flat connections
\begin{equation}\label{4.12}
\CA=\phi_S^{-1}\dd\phi_S\quad\mbox{with}\quad\phi_S\in\CG_S
\end{equation}
are parametrized by the infinite-dimensional group $\CG_S/G$. Action of $\CG_S$ maps them one into another.

\smallskip

\noindent
{\bf Extended phase space.} The framing of the bundle $E$ over $\R\times S$ reduces the gauge group $\CG$ to the subgroup $\CG_0$ in \eqref{4.5} and extends the phase space \eqref{3.16} to \eqref{4.11} since gluons become massive in the region $S$. On the other hand, framing means introducing additional degrees of freedom described by data \eqref{4.4}. At this point, we can already guess that the field $\phi_S\in\CG_S$ is a local version of the Stueckelberg field (supported on $S$) and therefore full $\CG$-invariance can be restored on the extended phase space
\begin{equation}\label{4.13}
T^*\Abb^{}_{\Sigma} \times T^*\CG^{}_{S}\cong T^*(\Abb^{}_{\Sigma} \times \CG^{}_{S})
\end{equation}
with $(\CA , \CE )\in  T^*\Abb^{}_{\Sigma}$ and $(\phi_S,\pi_S)\in T^*\CG^{}_{S}$, 
where $\pi_S\in (\mbox{Lie}\,\CG^{}_{S})^*\cong \mbox{Lie}\,\CG^{}_{S}$ is a ``momentum'' for $\phi_S$.

The group $\CG_S$ acts on $(\CA, \CE, \phi_S, \pi_S)$ from the extended phase space \eqref{4.13} by the formulae
\begin{equation}\label{4.14}
\CA_a\mapsto\CA_a^g=g^{-1}_S\CA_a g_S + g^{-1}_S\dpar_a g_S\ ,\quad \phi_S\mapsto \phi^g_S=\phi_Sg_S\ ,
\end{equation}
\begin{equation}\label{4.15}
\CE_a\mapsto\CE_a^g=g^{-1}_S\CE_a g_S\quad\mbox{and}\quad \pi_S\mapsto\pi_S^g=g^{-1}_S\pi_S g_S\ .
\end{equation}
Now we can consider the whole group $\CG=\CG_0\rtimes\CG_S$ as the group of gauge transformations and we can always reduce the space \eqref{4.13} by the action of the group $\CG$ similar to \eqref{3.16}: 
$\CM_{\tt phs}^{\tt framed} = T^*\Abb^{}_{\Sigma}\times T^*\CG^{}_{S}/\!/\CG$ (symplectic reduction). 
We will  discuss the proper Gauss law constraint in the next section. Notice that we can fix a gauge by choosing 
$g_S=\phi_S^{-1}$ in \eqref{4.14} so that $\phi^g_S=\unit_n$. In the next section we will see that in this gauge $\pi^g_S$ 
will be proportional to non-dynamical component $\CA_t$ and can be eliminated. 

Summing up preliminary results, we can say that
\begin{itemize}
\item 
$\phi_S$ is a $G$-valued Stueckelberg field naturally appearing in gauge theory for bundles $E$ framed over a compact submanifold $S$ of $\R^3$,
\item
$\phi_S$ is a compactly supported field parametrizing longitudinal components of gluons,
\item
gauge theory for connections on framed bundles $(E, \phi_S)$ is Yang-Mills-Stueckelberg theory with the field $\phi_S$ nontrivial only on a compact subspace $S$ of $\R^3$.
\end{itemize}

\smallskip

\noindent
{\bf Submanifold $S$.} The data \eqref{4.4} defining gauge theories on framed bundles $(E, \phi_S)$ also contains the manifold $S$ embedded into $\Sigma =\R^3$. We will choose $S$ as a 3-dimensional submanifold of $\R^3$ diffeomorphic to the closed 3-ball $\bar B^3_R(0)$ of radius $R$ centered at $x=0$,
\begin{equation}\label{4.16}
S_0 = \bar B^3_R(0) = \{x\in\R^3\mid r^2=(x^1)^2+(x^2)^2 + (x^3)^2\le R^2\}\ .
\end{equation}
So, consider a smooth map
\begin{equation}\label{4.17}
X: \ \bar B^3_R(0)\to \R^3
\end{equation}
defined by three smooth functions $X^a=X^a(x^b)$ for $x^b\in \bar B^3_R(0)$. We introduce $S\subset\R^3$ as an embedded regular submanifold
\begin{equation}\label{4.18}
S=X(S_0)\subset \R^3
\end{equation}
diffeomorphic to the 3-ball $S_0=\bar B^3_R(0)$. We consider a proper isometric embedding  \eqref{4.18} such that $S$ is smooth, closed and bounded.

\smallskip

\noindent
{\bf Bump functions.} Now we should specify functions supported on $S$. Recall that a {\it characteristic function} of a subset $S$ in $\Sigma$ is the function
\begin{equation}\label{4.19}
\unit_S:\ \Sigma\to \{0,1\}\ ,\quad 
\unit_S(x)=\left\{\begin{array}{l}1\ \mbox{for}\ x\in S\\0\ \mbox{for}\ x\not\in S\end{array}\right.
\end{equation}
The function $\unit_S$ indicates whether $x\in\Sigma$ belongs to $S$ or not. Obviously, $1{-}\unit_S$ is the characteristic function of the complement $\Sigma\backslash S$ of $S$ in $\Sigma$. 

Let $\dpar S\cong S^2$ be the boundary of $S$. Then one can define the inward normal derivative $\delta^{}_{S^2}(x)$ of $\unit_S$ that is a surface delta function generalizing the Dirac delta function. To avoid the complexities associated with the use of generalized functions, one often uses {\it bump functions} which are smooth versions of characteristic functions. They are functions $f: \Sigma\to \R$ which are both smooth and compactly supported on $S\subset\Sigma$ and denoted $f_S$. These functions vanish outside $S$ similar to the characteristic function $\unit_S$. Bump functions $g_S: \Sigma\to G$ are unity outside $S$.

Let $\chi_S$ be a bump function supported on $S\subset \Sigma$ and $\xi$ is a $\frg$-valued function on $M$. Then the function 
\begin{equation}\label{4.20}
\xi_S=\chi_S\xi
\end{equation}
will be supported on $S$ for any $t\in\R$. It smooths out the function $\unit_S\xi$. The $G$-valued function
\begin{equation}\label{4.21}
g_S=\exp\xi_S
\end{equation}
is supported on $S$ and smooths out the function $\exp(\unit^{}_S\xi )$. Bump functions exist for $S\subset\Sigma$. 

\smallskip

\noindent
{\bf Examples.} To give an example of a bump function, we consider the space $\R^3$ and the ball $S_0$ in \eqref{4.16}. A bump function 
$\chi^{}_{S_0}$: $\R^3\to \R$ can be defined as
\begin{equation}\label{4.22}
\chi^{}_{S_0}=\left\{
\begin{array}{r}\exp\left (\frac{r^2}{r^2-R^2}\right )\ \mbox{for}\ r^2<R^2\\
0\qquad\qquad\mbox{for}\ r^2\ge R^2\end{array}\right .
\end{equation}
The function \eqref{4.22} can be written as
\begin{equation}\label{4.23}
\chi^{}_{S_0}=\exp\left (\frac{r^2}{r^2-R^2}\right )\, \unit^{}_{\{r<R\}}\ ,
\end{equation}
where $\unit^{}_{\{r<R\}}$ is the characteristic function of the open ball $B^3_R(0)$. Notice that instead of the ball $B^3_R(0)$ one can consider the cylinder 
\begin{equation}\label{4.24}
C_0=\{x\in\R^3\mid (x^1)^2+(x^2)^2<R^2, -L<x^3<L\}
\end{equation}
with the characteristic function $\unit^{}_{C_0}=\unit^{}_{B^2_R(0)}\cdot\unit^{}_{\{|x^3|<L\}} $ and the bump function supported on $C_0$. Other geometries of subsets $U_0\subset\R^3$ are also possible. 

\section{Yang-Mills-Stueckelberg theory with compactly supported fields}

\noindent
{\bf Action for boundary fields.} After introducing the additional dynamical variable $\phi_S$ supported on $S\subset\Sigma$, we should propose an action functional for it. We argued in Section 4 that $\phi^{}_S$ is the Stueckelberg field. Therefore, we can write action for it without difficulty. However, first we deviate aside and discuss the actions considered in the literature for edge modes $\vph^{}_{\dpar M}$, which are the closest relatives of the field $\phi_S$.

As far as I know, an action for additional dynamical variables $\vph^{}_{\dpar M}$ defined on the boundary $\dpar M=\R\times S^2_\infty$ of $M=\R^{3,1}$ was first discussed in the paper~\cite{GSW}. It was proposed the action
\begin{equation}\label{5.1}
S^0_{\dpar M} = \int_{\dpar M}\tr\{(\CA^{}_{|\dpar M}-\vph^{-1}_{\dpar M}\dd\vph^{}_{\dpar M})\wedge\ast^{}_3 j^{}_{\dpar M}\}\ ,
\end{equation}
where $\vph^{}_{\dpar M}$ is a $G$-valued field on $\dpar M$, $\ast^{}_3$ is the Hodge star operator on $\dpar M$ and $j^{}_{\dpar M}$ is a {\it dynamical} one-form on $\dpar M$. Variation w.r.t. $j^{}_{\dpar M}$ yields
\begin{equation}\label{5.2}
\CA^{}_{|\dpar M}= \vph^{-1}_{\dpar M}\dd\vph^{}_{\dpar M}\quad\mbox{on}\quad\dpar M=\R\times S^2_\infty\ .
\end{equation}
There are also equations (15) and (13) in \cite{GSW}  which say that 
\begin{equation}\label{5.3}
\ast^{}_4\CF^{}_{|\dpar M}=\ast^{}_3 j^{}_{\dpar M}\quad\mbox{and}\quad\dd^{}_{\CA}(\ast^{}_3j^{}_{\dpar M})=0\ .
\end{equation}
The action  \eqref{5.1} was further analyzed e.g. in~\cite{DF, BMV}.

Flatness \eqref{5.2} of $\CA$ on $\dpar M$ is natural at infinity but this may not be appropriate for boundaries at finite distance. This can be avoided if the current $j^{}_{\dpar M}$ is not dynamical but this assumption has many hidden flaws. Flatness \eqref{5.2} can also be avoided if one introduces the standard second order Lagrangian for  $\vph^{}_{\dpar M}$~\cite{MSTW} but then the component of $\CF$ orthogonal to $\dpar M$ in \eqref{5.3} will be proportional to $\CA^{}_{|\dpar M}$ what looks strange. All this can be avoided by considering the action
\begin{equation}\label{5.4a}
S^{}_{\dpar M} = \int_{\dpar M}\tr\left\{\CF^{}_{|\dpar M}\wedge\ast^{}_3\CF^{}_{|\dpar M} +
2 m^2 (\CA^{}_{|\dpar M}{-}\vph^{-1}_{\dpar M}\dd\vph^{}_{\dpar M})\wedge\ast^{}_3(\CA^{}_{|\dpar M}{-}\vph^{-1}_{\dpar M}\dd\vph^{}_{\dpar M})\right\},
\end{equation}
where $m^2$ is a constant. Then boundary field equations following from $S^{}_{\dpar M}$ and \eqref{3.10} will be
\begin{equation}\label{5.4}
\dd^{}_{\CA}(\ast^{}_3\CF^{}_{|\dpar M}) + \ast^{}_4\CF^{}_{|\dpar M}=\ast^{}_3j^{}_{\dpar M}\quad\mbox{and}\quad\dd^{}_{\CA}(\ast^{}_3j^{}_{\dpar M})=0
\end{equation}
for $j^{}_{\dpar M}=m^2(\CA^{}_{|\dpar M}-\vph^{-1}_{\dpar M}\dd\vph^{}_{\dpar M})$.
We will not discuss this action in this paper.

\smallskip

\noindent
{\bf Higgs field $\longrightarrow$ Stueckelberg field.}  The standard action for scalar fields is the Higgs action. The field $\phi_S\in G$ cannot vanish as it should be for a Higgs field for describing spontaneous symmetry breaking. That is why we consider first a field 
\begin{equation}\label{5.5}
\Phi_S\in\mbox{Mat}\,(n, \C)\ ,
\end{equation}
which is a complex $n{\times}n$ matrix supported on $S$. Consider the Lagrangian density
\begin{equation}\label{5.6}
\CL^{}_{\Phi_S} = -\frac{m^2_S}{2}\,\eta^{\mu\nu}\,\tr\left((\nabla_\mu\Phi^{}_S)^\+\nabla_\nu\Phi^{}_S)\right ) - \frac{1}{4}\ga_S\left((\sfrac1n\,\tr (\Phi_S^\+\Phi^{}_S) - 1\right)^2\ ,
\end{equation}
where
\begin{equation}\label{5.7}
\nabla^{}_\mu\Phi^{}_S=\dpar^{}_\mu\Phi^{}_S-\Phi^{}_S\CA^{}_\mu\ ,\quad 
m^{}_S=m\chi^{}_S\ ,\quad\ga^{}_S=\ga\chi^{}_S\ ,
\end{equation}
$m, \ga\ge 0$ are some real constants, and $\chi^{}_S$ is a bump function. The standard logic is that the minimum of energy is not at 
$\Phi^{}_S=0$ but at $\Phi^{}_S$ satisfying the equation
\begin{equation}\label{5.8}
\Phi_S^\+\Phi^{}_S=\unit_n\ .
\end{equation}
The framing field $\phi^{}_S\in G$ is the Higgs field satisfying \eqref{5.8} when the quartic term in the potential energy in \eqref{5.6} vanishes.

\smallskip

\noindent
{\bf Global breaking of symmetries.}  Suppose that the bundle $E$ over $M=\R\times\Sigma=\R\times\R^3=\R^{3,1}$ is framed
over the whole Minkowski space. This means that $S_0=\Sigma=\R^3$ and in the bump function \eqref{4.23} one should consider $R^2\to\infty$ and then $\chi^{}_{S_0}=1$ in the whole Minkowski space. Then $m^{}_S=m=\,$const and the Lagrangian density is
\begin{equation}\label{5.9}
\CL^{}_\phi = -\frac{m^2}{2}\,\eta^{\mu\nu}\,\tr(\nabla_\mu\phi)^\+\nabla_\nu\phi=  
\frac{m^2}{2}\,\eta^{\mu\nu}\,\tr(\CA_\mu-\phi^\+\dpar_\mu\phi)(\CA_\nu-\phi^\+\dpar_\nu\phi) \ ,
\end{equation}
where $\phi$ is the Stueckelberg field defined in the {\it whole} Minkowski space $\R^{3,1}$. Thus, we have the action functional
\begin{equation}\label{5.10}
S=\frac{1}{4}\int_{\R^{3,1}}\dd^4x \ \tr\left\{\CF_{\mu\nu}\CF^{\mu\nu} + 
2m^2(\CA_\mu-\phi^\+\dpar_\mu\phi)(\CA^\mu-\phi^\+\dpar^\mu\phi)\right\}
\end{equation}
of Yang-Mills-Stueckelberg theory describing massive gluon fields on $\R^{3,1}$ (see the review \cite{RuAl} and references therein). In Section 4 we have shown that such $\phi$ describes framing in the bundle $E$ over the whole space-time and generates the mass of gauge fields. This is easily seen in the gauge $\phi =\unit_n$ breaking the gauge symmetry.

\smallskip

\noindent
{\bf Local breaking of symmetries.}  For $\phi^{}_S\in G\subset\,$SU$(n)$ supported on $S\subset\R^3$, we obtain from  \eqref{5.6} the Lagrangian density
\begin{equation}\label{5.11}
\CL^{}_{\phi^{}_S} = -\frac{m^2_S}{2}\ \tr(\nabla_\mu\phi^{}_S)^\+\nabla^\mu\phi^{}_S
= \frac{m^2_S}{2}\ \tr(\CA_\mu-\phi^\+_S\dpar_\mu\phi_S)(\CA^\mu-\phi^\+_S\dpar^\mu\phi_S) \ ,
\end{equation}
describing framing of the bundle $E$ over $S$. In the gauge $\phi^{}_S=\unit_n$ we again obtain the mass term for gluons but it is nonzero only in the region $S$ since $m^{}_S=m\chi^{}_S$ with a bump function $\chi^{}_S$ supported on $S$. Thus, we have a local breaking of gauge symmetries only in the region $S\subset\R^3$.

Note that we consider $\phi^{}_S$ with values in the group $G$ which leads to the massiveness of {\it all} components $\CA^i$ of the gluon 
$\CA=\CA^iI_i$, where $\{I_i\}$ form the matrix basis of the Lie algebra $\frg$. However, one can choose $\phi^{}_S\in G/H$ for any closed subgroup $H$ in $G$, and then the components of $\CA$ with values in the subalgebra ${\frh} =\,$Lie$\,H$ will remain massless. This corresponds to a {\it partial} breaking of gauge symmetries in $S$. Technically, this can be done by choosing
\begin{equation}\label{5.12}
\phi^{}_S=g^{}_S\,T\, g^{\+}_S\ ,
\end{equation}
where $T$ is a coordinate-independent matrix from the Cartan subalgebra of $\frg$ and $g^{}_S\in \CG^{}_S$. Then $\phi^{}_S$ in \eqref{5.12} is an element of adjoint orbit $G/H$ of $G$ acting on $\frg \cong \frg^*$. One can easily check that after substitution of \eqref{5.12} into \eqref{5.11} the components of $\CA$ from the subalgebra $\frh\subset\frg$ will disappear from the Lagrangian density \eqref{5.11}.

\smallskip

\noindent
{\bf Dirichlet energy.} Let us go back to the data \eqref{4.4} defining framing. We see from \eqref{4.4} that we must start by specifying the submanifold               
 $S$ in $\R^3$. In \eqref{4.16}-\eqref{4.18} this submanifold was defined by three functions
\begin{equation}\label{5.13}
X^a:\quad S_0\ \to\ S\subset\R^3
\end{equation}
mapping the 3-ball \eqref{4.16} into $\R^3$. For simplification we consider $X^a$ as time-independent functions and add to them $X^0:=t$. These additional degrees of freedom can be described by the sigma model Lagrangian density
\begin{equation}\label{5.14}
\CL^{}_{X^{}_S} = -\frac{1}{2}\Lambda^{}_S\eta^{\mu\nu}\eta_{\lambda\sigma}\dpar_\mu X^\lambda\dpar_\nu X^\sigma\ ,
\end{equation}
where $\Lambda^{}_S=\Lambda\chi^{}_S$ with a constant $\Lambda >0$ and a bump function $\chi^{}_S$. For the static case, \eqref{5.14} will give the Dirichlet energy in the Hamiltonian. It is non-negative for any functions $X^a$ on the 3-ball $S_0$.

Variation of \eqref{5.14} gives the Laplace equation
\begin{equation}\label{5.15}
\Delta X^a=0\quad\mbox{on}\quad S_0=\bar B^3_R(0)
\end{equation}
with solutions
\begin{equation}\label{5.16}
X^a_l=R\left(\frac{r}{R}\right)^l f^a_m Y_{lm}\ ,
\end{equation}
where $f^a_m$ are constants and $Y_{lm}(\theta , \vph )$ are spherical harmonics on $\dpar S_0$. The lowest energy mode is given by
\begin{equation}\label{5.17}
l=1\ :\quad X^a_1=x^a\ ,
\end{equation}
describing the identity map $X: S_0\to S=S_0$. For \eqref{5.17} we have
\begin{equation}\label{5.18}
\CL^{}_{X^{}_S}= -2\Lambda\chi^{}_S\ ,
\end{equation}
where $\chi^{}_S$ is the bump function \eqref{4.22}. Higher harmonics \eqref{5.16} with $l\ge 2$ will simply increase the energy calculated from \eqref{5.14} by integrating over the 3-ball $S$ and it is reasonable to restrict oneself to the identy map \eqref{5.17} of $S_0$ into $\R^3$ with minimal energy. From a physical point of view, this is the energy required to create a bubble of the Stueckelberg vacuum inside the ordinary vacuum of Yang-Mills theory.

\smallskip

\noindent
{\bf Action and field equations.} Summarizing all the above, the action for gauge fields on the bundle $E\to\R^{3,1}$ framed over a submanifold $S\subset\R^3$ reads
\begin{equation}\label{5.19}
S=\int_{\R^{3,1}}\dd^4x\left\{\sfrac14\tr\left(\CF_{\mu\nu}\CF^{\mu\nu} + 
2m^2_S(\CA_\mu-\phi^\+_S\dpar_\mu\phi^{}_S)(\CA^\mu-\phi^\+_S\dpar^\mu\phi^{}_S)\right)-2\Lambda\chi^{}_S \right\}\ ,
\end{equation}
where we substituted \eqref{5.18} instead of \eqref{5.14}. The Euler-Lagrange equations following from \eqref{5.19} have the form
\begin{equation}\label{5.20}
\nabla_\mu\CF^{\mu\nu}= J^\nu_S\ ,
\end{equation}
\begin{equation}\label{5.21}
\eta^{\mu\nu}\nabla_\mu\nabla_\nu\phi^{}_S=0\quad\Leftrightarrow\quad\nabla_\mu  J^\mu_S=0\ ,
\end{equation}
where 
\begin{equation}\label{5.22}
J^\mu_S:=m^{2}_S(\CA^\mu - \phi^\+_S\dpar^\mu\phi^{}_S)\ ,\quad
\nabla_\mu\phi^{}_S:=\dpar_\mu\phi^{}_S - \phi^{}_SA_\mu\quad\mbox{and}\quad 
\nabla_\mu J^{\nu}_S:=\dpar_\mu J^{\nu}_S + [A_\mu , J^{\nu}_S].
\end{equation}
Note that the field equation for $\phi^{}_S$ is the equation of covariant constancy of the current $J^{\mu}_S$. On the other hand, this equation is the generalized Lorenz gauge condition
\begin{equation}\label{5.23}
\dpar_\mu (m^{2}_S\CA^\mu )=\nabla_\mu (m^2_S\phi^{\+}_S\dpar^\mu\phi^{}_S)
\end{equation}
of Yang-Mills-Stueckelberg theory. In the gauge $\phi^{}_S=\unit_n$ we get the Lorenz gauge condition
\begin{equation}\label{5.24}
\dpar_\mu (m^2_S\CA^\mu ) =0
\end{equation}
defining three independent components of gluons $\CA$ having mass $m^{}_S$ in the region $S\subset\R^3$.

Note that the temporal gauge $\CA_t=0$ in the massive case is not possible but nevertheless the field $\CA_t=\CA_0$ is not dynamical. It can be eliminated by using the $\nu =0$ component in \eqref{5.20} and in the Hamiltonian formalism we will have only the fields
\begin{equation}\label{5.25}
(\CA_a, \CE_a, \phi^{}_S, \pi^{}_S) ,
\end{equation}
where $\pi_S=\dpar\CL/\dpar\dot\phi^{}_S$. The Hamiltonian density for the model \eqref{5.19} is
\begin{equation}\label{5.26}
\CH = 2\Lambda_S + \sfrac12(\CE^i_a\CE^i_a + \CB^i_a\CB^i_a + \tilde J^i_t\tilde J^i_t + \tilde J^i_a\tilde J^i_a)\quad\mbox{for}\quad 
\tilde  J^\mu= \frac{1}{m_S}J^\mu_S
\end{equation}
and it is obvious that $\CH\ge 2\Lambda^{}_S$. Using \eqref{5.14} simply changes the first term in \eqref{5.26} by extra positive terms (the Dirichlet energy density). Integral of $\CH_0 = 2\Lambda_S$ over the 3-ball $S_0=\bar B^3_R(0)$ is proportional to 
\begin{equation}\label{5.27}
H_0=\Lambda R^3
\end{equation}
and grows with increasing $R$.

\smallskip

\noindent
{\bf Quarks in a bubble of Stueckelberg vacuum.} Local breaking of gauge symmetry described in this paper makes it possible to formulate a new confinement scenario as follows. Placing quarks in a 3-ball $S_0$ of small radius $R$ fixes the frame in the bundle $E\to\R^{3,1}$ and leads to the condensation of gluons in the Stueckelberg field $\phi^{}_S\in G$ supported on $S_0$. The creation of a Stueckelberg vacuum bubble $S_0$ requires at least the Dirichlet energy \eqref{5.27}. This energy has the same form as in the MIT bag model and provides an attraction of quarks. On the other hand, kinetic energy of moving quarks leads to a repulsion, so that there is an equilibrium at some $R_0$. Thus, appearing of a bubble of Stueckelberg vacuum around quarks can be responsible for the confinement. The field $\phi^{}_S$ is contained implicitly in QCD (longitudinal component of gluons) and does not require its introduction from the outside.

If we accept the above picture of symmetry breaking, then asymptotic freedom can also be described in terms of compactly supported functions. Namely, let us replace $\CA$ in the action \eqref{5.19} as follows
\begin{equation}\label{5.29}
\CA = (1-\zeta^{}_S) \tilde\CA
\end{equation}
where $\zeta^{}_S$ is a function supported on $S$, e.g. the bump function $\chi^{}_S$. Then outside $S$ we have $\CA=\tilde\CA$ and for $r\to 0$ we have $\chi^{}_S\to 1$, i.e. the effective gauge coupling 
\begin{equation}\label{5.30}
g^{\tt eff}_{\tt YM} = (1-\chi^{}_S)g_{\tt YM}\to 0\ .
\end{equation}
Note that Fourier transform of $\chi^{}_S$ given in \eqref{4.23} is a real analytic function which decays asymptotically to zero for large momenta. Thus, not only $m^{}_S$ and $\Lambda^{}_S$ but also the gauge coupling $g^{\tt eff}_{\tt YM}$ can vary in $S\subset\R^3$.

\section{Framing in gravity}

The topic of this paper is framing in gauge theories, so we will only briefly touch on the issues of framing in gravity. 

\smallskip

\noindent
{\bf Tetrads.} Consider a four-dimensional Lorentzian manifold $M$ with a metric 
\begin{equation}\label{6.1}
\dd s^2=g_{\mu\nu}\dd x^\mu\dd x^\nu
\end{equation}
in local coordinates $x^\mu\in U\subset M$, $\mu ,\nu = 0,...,3$. Let $TM$ be the tangent bundle of $M$ and $F(TM)=P(M,$\,SO(3,1)) the associated bundle of orthonormal frames on $TM$. Four smooth vector fields (a tetrad or vierbein) 
\begin{equation}\label{6.2}
e_a = e_a^\mu\dpar_\mu\quad\mbox{for}\quad a=0,...,3\ ,
\end{equation}
define a frame on $TM$ over $U\subset M$. The local sections \eqref{6.2} of $TM$ are dual to one-forms (a co-tetrad or co-vierbein)
\begin{equation}\label{6.3}
e^a=e^a_\mu\dd x^\mu\ ,\quad e_a\lrcorner e^b = e_a^\mu e^b_\mu =\delta_a^b
\end{equation}
which define a frame on the cotangent bundle $T^*M$. In terms of the co-vierbein \eqref{6.3}, the metric \eqref{6.1}  reads
\begin{equation}\label{6.4}
\dd s^2=\eta_{ab}e^ae^b
\end{equation}
with $\eta = (\eta_{ab})=\,$diag$(-1, 1, 1, 1)$.

\smallskip

\noindent
{\bf Spin connection.} On the bundles $P(M,$  SO(3,1)), $TM$ and $T^*M$ we can define a spin connection $\omega_\mu =(\omega_{\mu\ b}^a)$ needed to introduce the covariant derivative of spinors on curved manifolds $M$. Relation between a torsion-free spin connection $\omega_\mu$ and (co-)vierbeins \eqref{6.2}-\eqref{6.3} appears after imposing the equations
\begin{equation}\label{6.5}
\nabla_\mu e^a_\nu =\dpar_\mu e^a_\nu + \omega_{\mu\ b}^a e^b_\nu - \Gamma^\sigma_{\mu\nu}e^a_\sigma =0\ ,
\end{equation}
where $\Gamma^\sigma_{\mu\nu}$ are the Christoffel symbols. Tetrads are used in the first order {\it Palatini formulation} of Einstein-Hilbert action in terms of tetrads and spin connection. 

\smallskip

\noindent
{\bf Changing of frame.} The frame $\{e_a\}$ is transformed under diffeomorphisms
\begin{equation}\label{6.6}
X: \ x^\mu \to X^\mu = X^\mu (x^\nu)\ \Rightarrow\ e^\mu_a \to \tilde e^\mu_a=\frac{\dpar X^\mu}{\dpar x^\nu}e^\nu_a\ ,
\end{equation}
where $X\in\,$Diff\,$M$. The frame $\{e_a\}$ is also transformed under local Lorentz rotations at any point $x\in M$,
\begin{equation}\label{6.7}
e_a \to \hat e_a =L_a^be_b\ ,
\end{equation}
where
\begin{equation}\label{6.8}
L=(L_a^b (x))\in \mbox{SO}_x(3,1)
\end{equation}
are local sections of the frame bundle $P(M, $ SO(3,1)) or associated bundle Int$\,P$.  Diffeomorphisms \eqref{6.6} and local Lorentz transformations \eqref{6.7}  define automorphisms of the bundle  $P(M, $ SO(3,1)) as well as the tangent bundle $TM$. Field equations of gravity plus matter (e.g. Einstein-Dirac equations) are considered invariant under these transformations.

\smallskip

\noindent
{\bf Local breaking of diffeomorphism invariance.} Suppose that frame $\{e_a\}$ on $TM$ is fixed over a submanifold $N$ of $M$, i.e. on the restriction $TM_{|N}$ of the tangent bundle $TM$ to $N\subset M$. Then the subgroup Diff$\,N$ of Diff$\,M$, defined by smooth maps
\begin{equation}\label{6.9}
X_N :\ N\to M\quad\Leftrightarrow\quad X_N :\ x^\mu \to X^\mu_N (x^\nu)\ ,\ x\in N,
\end{equation}
is broked. Let us consider diffeomorphisms which are identity on $N\subset M$ and denote the group of such diffeomorphisms by Diff$_0M$. Then Diff$_0M$ is a normal subgroup in Diff$\,M$ and we have the semidirect product of groups, 
\begin{equation}\label{6.10}
\mbox{Diff}\,M = \mbox{Diff}_0M\rtimes\mbox{Diff}\,N \ .
\end{equation}
The Diff$\,M$-invariance is restored if the gravity phase space is extended by the degrees of freedom \eqref{6.9} corresponding to Diff$\,N$. This can be done by adding a sigma model type action to the Einstein-Hilbert action, 
\begin{equation}\label{6.11}
S=\int_M\dd^4x \sqrt{-g}\left\{R[g] - \sfrac12\Lambda^{}_Ng^{\mu\nu}g_{\lambda\sigma}\dpar_\mu X^\lambda_N  \dpar_\nu X^\sigma_N  \right\}\ ,
\end{equation}
where
\begin{equation}\label{6.12}
\Lambda_N=\Lambda\chi^{}_N
\end{equation}
is a ``cosmological term''  supported on $N$. Here $\Lambda$ is constant and $\chi^{}_N$ is a real-valued function supported on $N$. Choosing $N=\R\times S$ with a compact spacial submanifold $S$ in $M$, we can consider 
\begin{equation}\label{6.13}
\Lambda^{}_S=\Lambda\chi^{}_S
\end{equation}
with a bump function $\chi^{}_S$. For identity diffeomorphism $X_N\in\,$Diff$\,N$ we have
\begin{equation}\label{6.14}
X_N^\mu = x^\mu\ .
\end{equation}
For this case we obtain in \eqref{6.11} a ``cosmological term'' supported on $N\subset M$ or on spacial $S$. Anyway, the terms \eqref{6.12} and \eqref{6.13} are related with the vacuum energy density which is nonzero only on a subspace of space-time $M$. Other choices of additional terms in \eqref{6.11} are possible. 


\noindent
{\bf Local breaking of Lorentz invariance.} If the frame $\{e_a\}$ on $TM$ is fixed over $N\subset M$ then the group $\CG^{\tt SO(3,1)}$ of local Lorentz transformations \eqref{6.7} is broken to a subgroup $\CG_0^{\tt SO(3,1)}$ of transformations which are identity on $N\subset M$. Then we have 
\begin{equation}\label{6.15}
\CG^{\tt SO(3,1)}=\CG_0^{\tt SO(3,1)}\rtimes \CG_N^{\tt SO(3,1)}
\end{equation}
since $\CG_0^{\tt SO(3,1)}$ is a normal subgroup in $\CG^{\tt SO(3,1)}$. Broken Lorentz symmetries can be restored by introducing SO(3,1)-valued field  $L_N(x)$ rotating $e_a$ as in \eqref{6.7} and by adding a Lagrangian for the field $L_N$. For instance, one can consider the term 
\begin{equation}\label{6.16}
\CL^{}_{L_N} = \chi^{}_N g^{\mu\nu} (\omega_{\mu\ b}^a - L_{Nc}^a\dpar_\mu L^c_{Nb})(\omega^b_{\nu\ a} - L_{Nd}^b\dpar_\nu L^d_{Na})
\end{equation}
analogous to the term \eqref{5.11} in gauge theory. Both $X_N=(X^\mu_N)$ and $L_N=(L^a_{Nb})$ are the well-known Stueckelberg fields in gravity.

\noindent
{\bf Local mass for gravitons.} Let us consider the term
\begin{align}\nonumber
S^{}_{\tt mass}= -\sfrac12\ m^2\veps_{abcd}\int_M&\chi^{2}_N(\al_0 E_N^{a}\wedge E_N^{b}\wedge E_N^{c}\wedge  E_N^{d}+\al_1F^a_N\wedge E_N^{b}\wedge E_N^{c}\wedge  E_N^{d}
\\
&+\al_2F^a_N\wedge F^b_N\wedge  E_N^{c}\wedge  E_N^{d} +\al_3F^a_N\wedge F^b_N\wedge F^c_N\wedge  E_N^{d} )\ ,
\label{6.17}\end{align}
where 
\begin{equation}\label{6.18}
F^a_N=\delta^a_\mu\frac{\dpar X^\mu_N}{\dpar x^\nu}\dd x^\nu\quad\mbox{and}\quad E_N^{a}=L_{Nb}^{a}e^b
\end{equation}
are supported on $N$. For $N=M$ the term \eqref{6.17} with $\chi^{}_N=1$ is used in gravity for describing massive graviton~\cite{OnTol}. The functions $\chi^{}_N, X^\mu_N$ and $L_{Nb}^{a}$ supported on $N$ change the theory since \eqref{6.17} leads to a mass $m^{}_N=m\chi^{}_N$ for the graviton only in the region $N\subset M$. For $N=\R\times S$ the spacial region $S$ can be small and the mass can appear only in the region $S$. Of course, there can be a lot of such regions $S_1, ..., S_l, ...$ in space.

The requirement for the massiveness of the graviton in the entire space-time $M$ may be too strong.  It is proposed to consider  breaking and restorations of diffeomorphism and Lorentz symmetries  with the help of Stueckelberg fields $X_N$ and $L_N$ in a region $N\subset M$ and, in particular, in compact regions $S_1, ..., S_l, ...$ of space. This could provide a clue to understanding many as yet unexplained phenomena.

\section{Conclusions}

In recent years, there has been a lot of activity in the study of asymptotic/boundary symmetries, auxiliary edge/boundary fields, soft theorems and related topics. This paper aims to provide a general basis for these studies. Furthermore, we propose to consider not only boundaries of space-time $M$ but also submanifolds $N\subset M$ of any dimension dim$\,N=k+1\le 4$. In this paper, we have focused on spaces $M=\R\times\Sigma$ and $N=\R\times S$ with a three-dimensional compact spacial  submanifold $S$ of $\Sigma$. This was done to avoid blurring general ideas by the standard logic used when considering boundary terms. 

The general ideas of this paper are as follows. 

On the tangent bundle $TM$ of a curved manifold $M$ the symmetry groups are diffeomorphisms Diff$\,M$ mapping $M$ into itself and the group $\CG^{\tt SO(3,1)}$ of local Lorentz rotations of frames on $TM$. Together they define the group of automorphisms of the tangent bundle $TM$. If a vector $G$-bundle $E\to M$ is also given, then the group $\CG^G$ of automorphisms of the bundle $E$ is also given.

Suppose some submanifold $N$ of $M$ is given. For simplicity, we considered $N=\R\times S$, where $S$ is the spacial part of dimension $k=0,1,2,3$. We discussed the case $k=3$ and for the case of boundary one should consider $k=2$. After fixing $N$ in $M$ by using {\it adapted} coordinates one should define the group Diff$_0M$ which is identity on $N$. Then one should introduce the scalar Stueckelberg fields $X_N$ parametrizing the quotient group Diff$\,N$:=Diff$\,M$/Diff$_0M$.

For the group $\CG^{\tt SO(3,1)}$ of Lorentz transformations one should consider the subbundle $TN$ of $TM_{|N}$ and define the group $\CG_0^{\tt SO(3,1)}$ preserving a fixed frame on $TN$. Then one should introduce SO(3,1)-valued Stueckelberg field $L_N$ parametrizing the quotient group $\CG_N^{\tt SO(3,1)}:=\CG^{\tt SO(3,1)}/\CG_0^{\tt SO(3,1)}$. For a gauge vector bundle $E$ framed over $N\subset M$, one introduces the $G$-valued Stueckelberg field $\phi^{}_N$ discussed in detail in this paper. 

All Stueckelberg's fields $\phi^{}_N$, $X_N$ and $L_N$ are part $\Gla$ of the group $\CG$ of automorphisms of the model in question, which were transferred from the status of redundant (small/proper) symmetries to the status of physical (large/improper) symmetries due to framing over a submanifold $N$ in $M$.

It is necessary to set the action functionals for the fields $\phi^{}_N$, $X^{}_N$ and $L^{}_N$. Only based on the type of action, one can conclude whether fields $\phi^{}_N$, $X^{}_N$ and $L^{}_N$ are Stueckelberg fields or some other fields of an unclear nature. In this paper, we wrote out such actions and argued that $\phi^{}_N$, $X^{}_N$ and $L^{}_N$ are Stueckelberg fields. We considered the case $N=\R\times S$ with a compact space $S$ and showed that the condensation of a part of the automorphisms into the Stueckelberg fields $\phi^{}_S(t)$, $X^{}_S(t)$ and $L^{}_S(t)$ changes the properties of the vacuum in the region $S$. In particular, this vacuum has a positive energy density function  $\Lambda^{}_S$ that has a compact support and is responsible for color confinement.

The emergence of bubbles of Stueckelberg vacua can help to explain dark energy and many other as yet incomprehensible phenomena. 
Another interesting question is related to the Stueckelberg vacuum bubble in a small region $S$ around a large gravitating mass. It would be interesting to know if the Penrose-Hawking singularity theorems are preserved in the presence of the Stueckelberg fields $X^{}_N$ and $L^{}_N$. This requires additional study.

\bigskip

\noindent {\bf Acknowledgments}

\noindent
This work was supported by the Deutsche Forschungsgemeinschaft grant LE~838/19.


\bigskip

\begin{thebibliography}{99}

\bibitem{GSW}
J.L.~Gervais, B.~Sakita and S.~Wadia,
``The surface term in gauge theories,''\\
Phys. Lett. B \textbf{63} (1976) 55.

\bibitem{BCT}
R.~Benguria, P.~Cordero and C.~Teitelboim,
``Aspects of the Hamiltonian dynamics of interacting gravitational gauge and Higgs fields with applications to spherical symmetry,''\\
Nucl. Phys. B \textbf{122} (1977) 61.

\bibitem{Str1}
A.~Strominger,
``Asymptotic symmetries of Yang-Mills theory,''\\
JHEP \textbf{07} (2014) 151
[arXiv:1308.0589 [hep-th]].

\bibitem{DF}
W.~Donnelly and L.~Freidel,
``Local subsystems in gauge theory and gravity,''\\
JHEP \textbf{09} (2016) 102
[arXiv:1601.04744 [hep-th]].

\bibitem{BMV}
A.~Blommaert, T.G.~Mertens and H.~Verschelde,
``Edge dynamics from the path integral --- Maxwell and Yang-Mills,''
JHEP \textbf{11} (2018) 080
[arXiv:1804.07585 [hep-th]].

\bibitem{Str2}
A.~Strominger,
{\it Lectures on the infrared structure of gravity and gauge theory,}\\
Princeton University Press, Prinston, 2018.

\bibitem{Do1}
S.K.~Donaldson,
``Boundary value problems for Yang-Mills fields,''\\
J. Geom. Phys. \textbf{8} (1992) 89.

\bibitem{RT}
T.~Regge and C.~Teitelboim,
``Role of surface integrals in the Hamiltonian formulation of general relativity,''
Ann. Phys. \textbf{88} (1974) 286.

\bibitem{Do2}
S.K.~Donaldson,
``Instantons and geometric invariant theory,''\\
Commun. Math. Phys. \textbf{93} (1984) 453.

\bibitem{Lu}
M.~L\"ubke, ``The analytic moduli space of framed vector bundles,'' \\
J.~reine angew.~Math. {\bf 441} (1993) 45.

\bibitem{FS}
L.D.~Faddeev and A.A.~Slavnov, 
{\it Gauge fields: Introduction to quantum theory},\\ 
Benjamin/Cummings, Reading, MA, 1980.

\bibitem{MSTW}
P.~Mathieu, L.~Murray, A.~Schenkel and N.J.~Teh,
``Homological perspective on edge modes in linear Yang-Mills and Chern-Simons theory,''
Lett. Math. Phys. \textbf{110} (2020) 1559
[arXiv:1907.10651 [hep-th]].

\bibitem{RuAl}
H.~Ruegg and M.~Ruiz-Altaba,
``The Stueckelberg field,''\\
Int. J. Mod. Phys. A \textbf{19} (2004) 3265
[arXiv:hep-th/0304245 [hep-th]].

\bibitem{OnTol}
N.A.~Ondo and A.J.~Tolley,
``Complete decoupling limit of ghost-free massive gravity,''\\
JHEP \textbf{11} (2013) 059
[arXiv:1307.4769 [hep-th]].

\end{thebibliography}
\end{document}